\pdfoutput=1
\documentclass[sigconf]{acmart}

\AtBeginDocument{%
  \providecommand\BibTeX{{%
    \normalfont B\kern-0.5em{\scshape i\kern-0.25em b}\kern-0.8em\TeX}}}

\copyrightyear{2022}
\acmYear{2022}
\setcopyright{rightsretained}
\acmConference[UIST '22]{The 35th Annual ACM Symposium on User Interface Software and Technology}{October 29-November 2, 2022}{Bend, OR, USA}
\acmBooktitle{The 35th Annual ACM Symposium on User Interface Software and Technology (UIST '22), October 29-November 2, 2022, Bend, OR, USA}
\acmDOI{10.1145/3526113.3545703}
\acmISBN{978-1-4503-9320-1/22/10}

\usepackage{multirow}
\usepackage{soul}
\usepackage{xcolor}
\usepackage{amsmath}
\usepackage{booktabs}
\usepackage{graphicx}

\begin{document}



\title{CrossA11y: Identifying Video Accessibility Issues \\ via Cross-modal Grounding}


\author{Xingyu ``Bruce'' Liu}
\affiliation{%
  \institution{UCLA}
  \city{Los Angeles}
  \country{USA}}
 \email{xingyuliu@ucla.edu}

\author{Ruolin Wang}
\affiliation{%
  \institution{UCLA}
  \city{Los Angeles}
  \country{USA}}
 \email{violynne@ucla.edu}

\author{Dingzeyu Li}
\affiliation{%
  \institution{Adobe Research}
  \city{Seattle}
  \country{USA}}
 \email{dinli@adobe.com}

\author{Xiang `Anthony' Chen}
\affiliation{%
  \institution{UCLA}
  \city{Los Angeles}
  \country{USA}}
 \email{xac@ucla.edu}

\author{Amy Pavel}
\affiliation{%
  \institution{The University of Texas at Austin}
  \city{Austin}
  \country{USA}}
 \email{apavel@cs.utexas.edu}


\definecolor{darkgreen}{rgb}{0,0.5,0}
\definecolor{orange}{rgb}{1,0.5,0}
\definecolor{teal}{rgb}{0,0.5,0.5}
\definecolor{darkpurple}{rgb}{0.5, 0, 0.5}
\definecolor{burntorange}{rgb}{0.8, 0.3, 0}

\newcommand{\eg}{\textit{e.g., }}

\newcommand{\etal}{et al. }

\newcommand {\systemname}{CrossA11y}
\newcommand {\systemnamespace}{CrossA11y }

\begin{abstract}
Authors make their videos \textit{visually accessible} by adding audio descriptions (AD), and \textit{auditorily accessible} by adding closed captions (CC).
However, creating AD and CC is challenging and tedious, especially for non-professional describers and captioners, due to the difficulty of identifying accessibility problems in videos.
A video author will have to watch the video through and manually check for inaccessible information frame-by-frame, for both visual and auditory modalities.
In this paper, we present \systemname, a system that helps authors efficiently detect and address visual and auditory accessibility issues in videos.
Using cross-modal grounding analysis, \systemnamespace automatically measures accessibility of visual and audio segments in a video by checking for \textit{modality asymmetries}.
\systemnamespace then displays these segments and surfaces visual and audio accessibility issues in a unified interface, making it intuitive to locate, review, script AD/CC in-place, and preview the described and captioned video immediately.
We demonstrate the effectiveness of \systemnamespace through a lab study with 11 participants, comparing to existing baseline.


\end{abstract}




\begin{teaserfigure}
\centering
\includegraphics[width=7in]{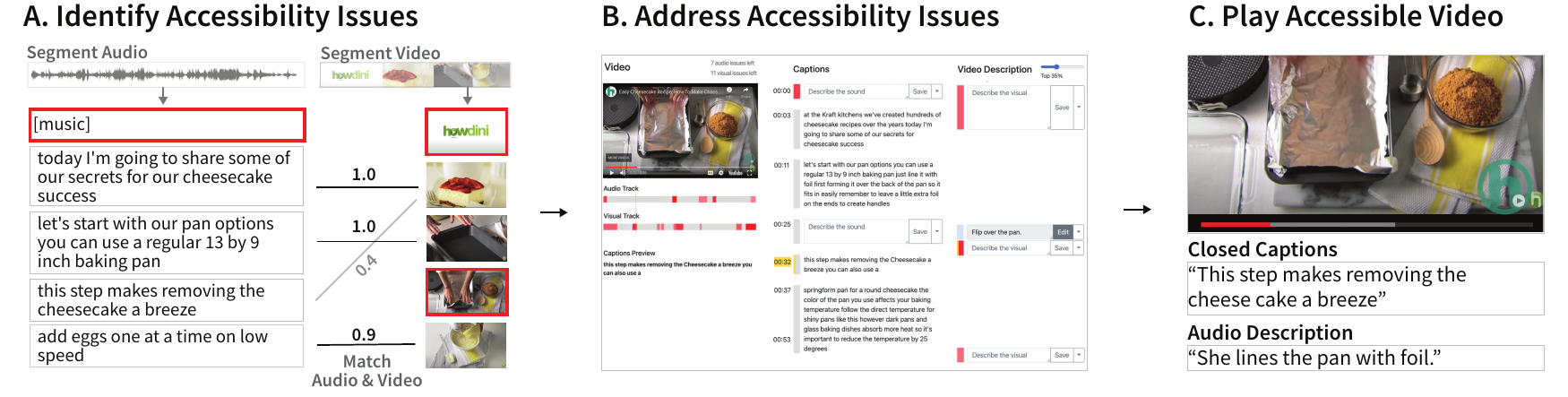}
\caption{Our system (A) identifies accessibility issues by locating \textit{modality asymmetries} (in red) between audio segments and video segments using cross-modal grounding. (B) lets authors address accessibility issues using the \systemnamespace interface to write captions and video descriptions, and (C) creates a more accessible video from the authored descriptions.}
\label{fig:system-diagram}
\end{teaserfigure}

\maketitle

\section{Introduction}

Videos use audio and visuals to convey information, making them inaccessible to people who cannot see or hear content in certain modalities. To make videos accessible, video authors add audio descriptions (AD) that describe important visual content, and closed captions (CC) that transcribe the speech and non-speech sounds. 
However, to identify parts of the video that require further description, authors must manually watch the video all the way through, playing and pausing to check if: (1) the important visuals have not been described in the audio (\eg~a travel montage in a vlog), and (2) the important audio is not present in the visuals and captions (\eg~a door slams off-screen).
This process of identifying inaccessible video segments is challenging and time-consuming, especially for 
video 
accessibility novices.

To guide authors to describe inaccessible video segments, existing audio description authoring tools surface ``gaps in speech'' as a proxy for moments where the visuals are unlikely to be verbally described~\cite{natalie2021efficacy, wang2021toward, campos2020cinead, pavel2020rescribe, yuksel2020human}.
However, many video genres including tutorials, vlogs, and lectures may not feature significant gaps in speech~\cite{liu2021makes, peng2021slidecho}, and audio description guidelines as well as prior research~\cite{wang2021toward, liu2021makes, wcag2} indicate that visuals can be inaccessible to blind and visually impaired (BVI) people even when there is accompanying speech. 
For example, a speaker may make an ambiguous verbal reference to visual content (\eg ``make sure to have \emph{these} before you get started'') or share a personal story while demonstrating a tutorial step. 
In addition, visuals without speech can be accessible if they are understandable from non-speech sounds alone.
Thus, using gaps in speech alone, authors will miss important inaccessible moments or be prompted to describe already-accessible moments. 
Similarly, caption authoring tools~\cite{lasecki2012real, rev, descript} let authors correct errors from automatic speech recognition (ASR), but they fail to surface moments when important audio does not also appear on screen (\eg someone leaves and we hear a door slam).

To help authors efficiently identify and address audio and visual accessibility problems, we present \systemname. \systemnamespace surfaces asymmetries between the visual track and the audio track, or \textit{modality asymmetries}. 
By identifying moments in the visuals that are not available in the audio, \systemnamespace surfaces moments that are not accessible to blind and visually impaired (BVI) audience members. Similarly, by identifying moments in the audio that are not available in the visuals, \systemnamespace surfaces moments that are not accessible to d/Deaf and Hard of Hearing (DHOH) audience.
To automatically identify modality asymmetries, \systemname's computational pipeline segments the audio and visual tracks and uses \textit{cross-modal grounding} to identify mismatches between the two tracks (Figure~\ref{fig:system-diagram}A). 
\systemnamespace then displays the results in an interface where authors can jointly author closed captions and audio descriptions by easily navigating to inaccessible moments (Figure~\ref{fig:system-diagram}B). Authors can then preview and export their resulting audio descriptions and closed captions (Figure~\ref{fig:system-diagram}C).

We evaluated \systemnamespace in a user study with 11 video authors creating captions and audio descriptions for four videos. Authors more efficiently authored audio descriptions and captions with better precision and recall in addressing accessibility issues when using \systemname's modality asymmetry predictions than without these predictions. 
We also invited two video authors who frequently posted videos on YouTube to use \systemnamespace to make two of their own videos accessible, and reported that they would use \systemnamespace in their workflow to produce more accessible videos. 

In summary, we contribute:
\begin{itemize} 
    \item A pipeline to compute accessibility scores of the visual and audio segments of a video by checking for \textit{modality asymmetries} via cross-modal grounding.
    \item A unified tool that helps video authors to locate and address visual and auditory accessibility problems of a video. 
    \item A user study demonstrating that \systemnamespace improves people's efficiency and reduces their mental demand in identifying accessibility issues. 
\end{itemize}

\section{Related Work}

\subsection{Video Accessibility Guidelines}
The Web Content Accessibility Guidelines' (2.0) principle of Perceivable suggests that \textit{``Information and user interface components must be presentable to users in ways they can perceive''}~\cite{wcag2}.
Thus, authors need to make their videos perceivable to audiences who cannot see or hear the content by adding Closed Captions (CC) that use text to describe ``both speech and non-speech audio information needed to understand the media content''~\cite{wcag2} and Audio Descriptions (AD) that use text to describe ``important visual details that cannot be understood from the main soundtrack alone''~\cite{wcagad,packer2015overview}.

In prior work that interviewed blind and visually impaired YouTube audience, participants reported losing a sense of the video during visual content that was not well-described in the speech: ambiguous verbal references to visual content (\textit{e.g.}, look at `this'), unidentified sounds, undescribed text-on-screen, and others such as visual jokes~\cite{liu2021makes}.
Thus, we aim to help content creators identify and address moments that are visually inaccessible due to modality asymmetry. 

Unlike audio description guidelines that suggest narrating the ``important'' visual content~\cite{wcagad}, guidelines for Closed Captions by the Federal Communications Commission~\cite{fcc} require closed captions that describe spoken words and convey background noises and other sounds to the \textit{fullest extent possible} within a synchronous track. 
Thus, we help authors identify all moments where additional synchronous description is needed, and help authors prioritize \eg silent portions may not require further description.

\begin{figure*}[t]
\centering
\includegraphics[width=7in]{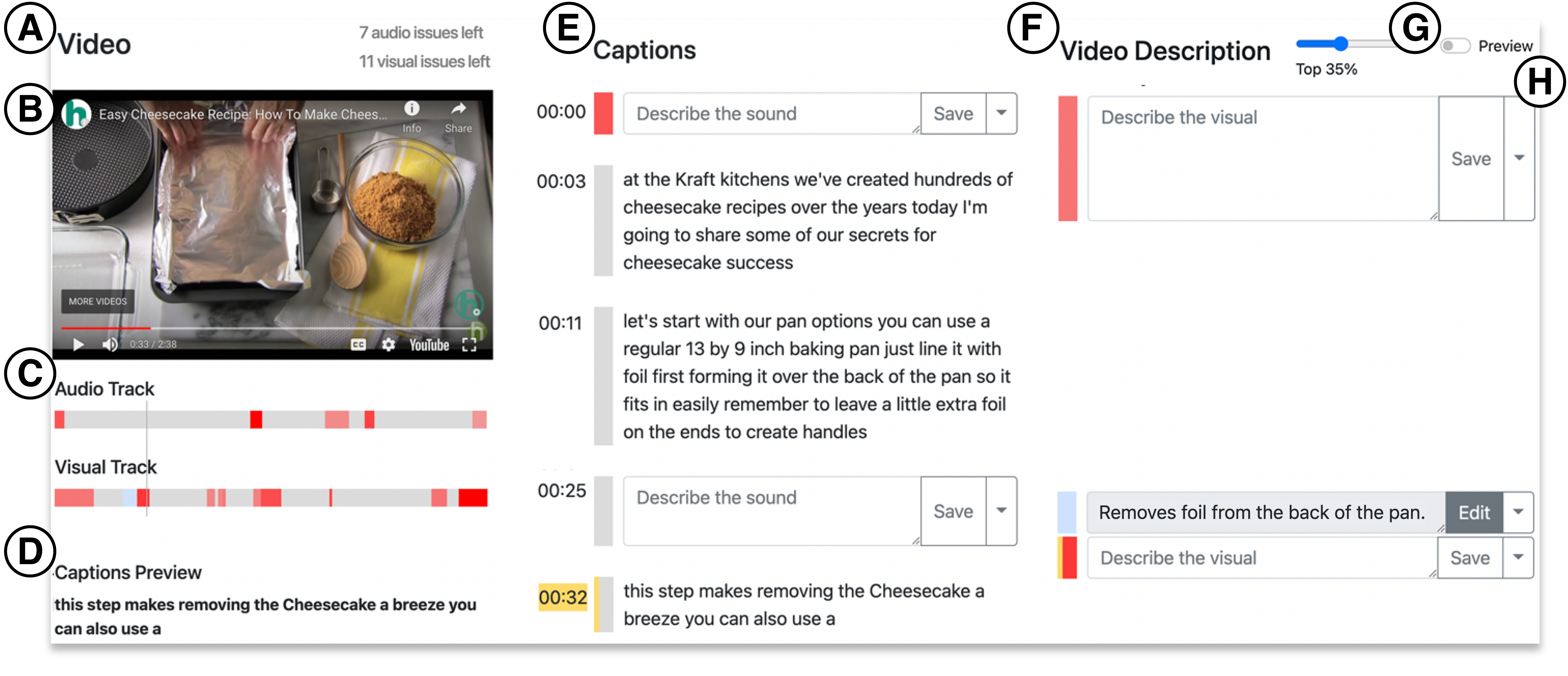}
\caption{In \systemname's interface, the \textit{video pane} (A) displays audio and visual timelines with accessibility visualization that allows authors to quickly identify and navigate to accessibility issues. The \textit{video description pane} (F) surfaces inaccessible visual segments and lets authors to add descriptions. The \textit{captions pane} (E) provides time-aligned captions and detected non-speech sound segments for authors to seek within the video and add captions. 
}
\label{fig:interface}
\end{figure*}

\subsection{Authoring AD and CC}
Prior work aims to help people manually author audio descriptions 
with task-specific authoring tools~\cite{youdescribe,branje2012livedescribe,3playmedia}, feedback on the content at production-time~\cite{peng2021say}, 
with feedback on audio descriptions~\cite{saray2011adaptive, natalie2021efficacy},
and with hosted descriptions~\cite{youdescribe}.
Since authoring descriptions is a time-consuming process, other prior work seeks to provide computational support for this task including using: computer vision to detect visual content~\cite{gagnon2009towards, gagnon2010computer, campos2020cinead}, using deep learning to provide a computer-drafted description~\cite{gagnon2009towards, yuksel2020human,wang2021toward}, synthesized voice to convert text to speech~\cite{gagnon2010computer, kobayashi2009providing, kobayashi2010synthesized,szarkowska2011text}, and automatic editing to fit human-authored descriptions into the space provided~\cite{pavel2020rescribe}. 
While focusing on methods to help people write better descriptions, such tools only find inaccessible moments for description by surfacing silent portions of the video~\cite{branje2012livedescribe,pavel2020rescribe,yuksel2020human,gagnon2010computer}, or by helping people find film-specific visual content that may need descriptions (\textit{e.g.}, scene changes, characters~\cite{gagnon2010computer}).
Rather than assessing video in a single modality, we explore finding accessibility problems by assessing asymmetries between the auditory and visual content.

Current caption authoring tools~\cite{lasecki2012real, rev, descript} transcribe speech and allow creators to correct the transcript. 
In this work, we also surface non-speech sounds to facilitate caption authoring and add modality matching score to help people prioritize points where additional description could be needed (\eg a sound that happens off-screen may be highly important to describe, while a silent section would not be).






\subsection{Accessibility Assessment Tools}
Assessing accessibility of visual content is challenging for authors who do not share accessibility needs with their audience members.
As a result, accessibility research includes a long history of prior work aimed to help people assess and correct accessibility problems in their designs including tools aimed at simulating accessibility issues~\cite{chromedevtools,asakawa2005s} and evaluating accessibility with respect to metrics~\cite{vigo2011automatic, mankoff2005your}.
Simulation-style tools to support sighted designers trying to achieve visually accessible designs; for example, Chrome Dev Tool's colorblindness and blurriness emulators to help designers assess legibility~\cite{chromedevtools}.
Using such simulations as a replacement for involvement with people with disabilities has several issues, as they are unable to capture the full experience of disability and give designers false conceptions~\cite{tigwell2021nuanced}. 
Given that people with disabilities, in partnership with organizations (\textit{e.g.}, W3C~\cite{wcag2}), have authored guidelines and best practices to make design accessible, other prior work alerts authors to violations of these guidelines in authoring tools. For example, accessibility checkers in PowerPoint~\cite{powerpointchecker} and Adobe Acrobat~\cite{acrobatchecker} alert authors to potential accessibility issues in their designs (\textit{e.g.}, missing alt text, document read order). 
Furthermore, web accessibility checkers provide a report card on similar types of issues to fix~\cite{vigo2011automatic, mankoff2005your}. 
We extend prior work by assessing the accessibility issues and surfacing accessibility issues based on existing guidelines about video accessibility.  

\subsection{Assessing Audio and Visual Similarity}
Recent work in unsupervised cross-modal machine learning explores learning a joint embedding space for information in different modalities, including text and images~\cite{clip2021learning, Wang2019CAMPCA, Stroud2020LearningVR}, text and video~\cite{mle2020end, mmv2020self, Croitoru2021TeachTextCG, Wang2021T2VLADGS}, and audio and video~\cite{mmv2020self, Morgado2021AudioVisualID}. 
Such models enable comparison between any visual, text, or audio segment.
While these models can be used for retrieval across modalities (\textit{e.g.}, text-image retrieval~\cite{clip2021learning, Wang2019CAMPCA}, and text-video retrieval~\cite{Croitoru2021TeachTextCG, Wang2021T2VLADGS}), we use a cross-modal approach to inform authors of accessibility issues due to low correspondence between the modalities, or modality asymmetry.

Prior video work in video accessibility has also considered the similarity between video and audio tracks. Wang et al. filter possible accessibility problems first by gaps in speech then use video and audio similarity to prioritize what non-speech segments to describe~\cite{wang2021toward}. Liu et al. checks if detected objects are mentioned in the transcript, along with other metrics, then assigns an accessibility score to a video to help blind viewers find accessible videos~\cite{liu2021makes}.  
We instead compute the fine-grained similarity between audio and visual segments to help authors find accessibility issues outside of gaps in speech that have not yet been addressed by prior work.

\section{\systemnamespace Interface}
\systemnamespace enables authors to efficiently identify and address visual and auditory accessibility problems in videos.
The interface consists of three main components: 1) a \textit{video pane} that lets authors navigate via an audio segment timeline or video segment timeline to identify inaccessible video segments (Figure~\ref{fig:interface}A), 2) the \textit{video description pane} that lets authors identify and address visual accessibility problems (Figure~\ref{fig:interface}F), and 3) a \textit{caption pane} that lets authors address auditory accessibility problems and navigate the video with a time-aligned caption transcript (Figure~\ref{fig:interface}E). 




\subsection{Video Pane}
\begin{figure}[t]
\centering
\includegraphics[width=0.7\linewidth]{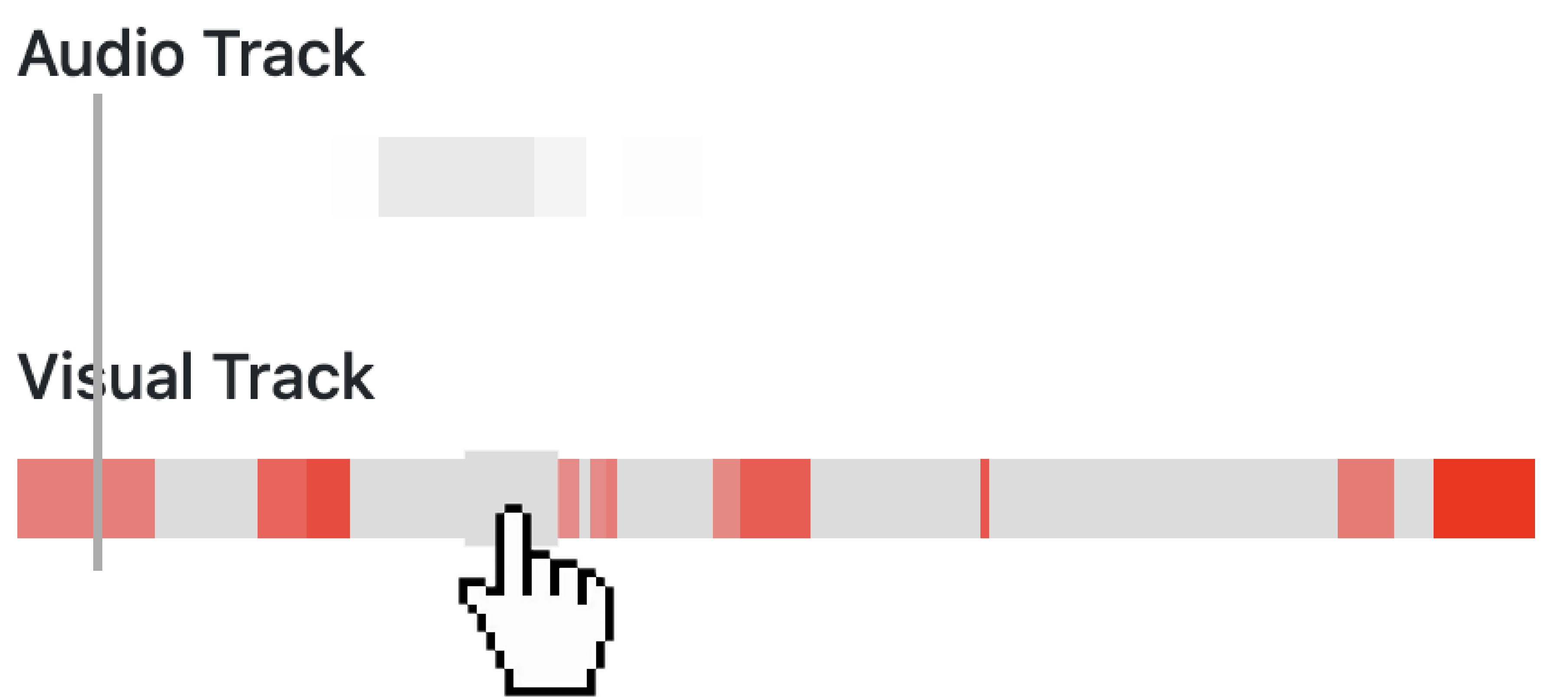}
\caption{Authors can hover on segments of visual/audio timelines in \systemnamespace to inspect the computed correspondence between the selected segment and segments in the other modality. Here, opaque segments in the audio track are the segments predicted by the system to match the segment in the video track.}
\label{fig:hover}
\end{figure}

The \textit{video pane} (Figure~\ref{fig:interface}A) displays the video and lets authors play/pause the video and seek within the video using two timelines: (1) the audio timeline that lets authors navigate to auditorily inaccessible segments, and (2) the visual timeline that lets authors navigate to visually inaccessible segments. 
The audio timeline displays audio segments that each represent a segment with continuous speech, or non-speech sound. The visual timeline displays visual segments that each represent a segment of continuous footage (i.e., a shot).
Each segment is colored with its estimated accessibility\footnote{Computed using the cross-modal grounding pipeline as described in Section~\ref{algorithmic_methods}} from gray (accessible) to red (inaccessible) using sRGB inverse gamma mixing.
The darkness of the red represents the weighted sum of the similarity score of a segment to segments in the other modality.
Using either timeline, authors can gain an overview of accessibility issues, or quickly navigate to an inaccessible segment by clicking on the segment to play the corresponding point in the video.
For example, by clicking on the first red segment in the audio track (Figure~\ref{fig:interface}C) an author will hear an inaccessible audio segment --- music plays that is not available in the captions preview (Figure~\ref{fig:interface}D). 
Authors may inspect inaccessibility prediction results displayed in the timeline by hovering over an audio or visual segment (Figure~\ref{fig:hover}) to see the audio segments that are predicted to match that segment (displayed with a higher opacity). 
As the author navigates and plays the video with the video pane, the corresponding segments are highlighted in the linked video description pane and the caption pane.

\subsection{Video Description Pane}
\label{video_description_pane}
\begin{figure}[t]
\centering
\includegraphics[width=\linewidth]{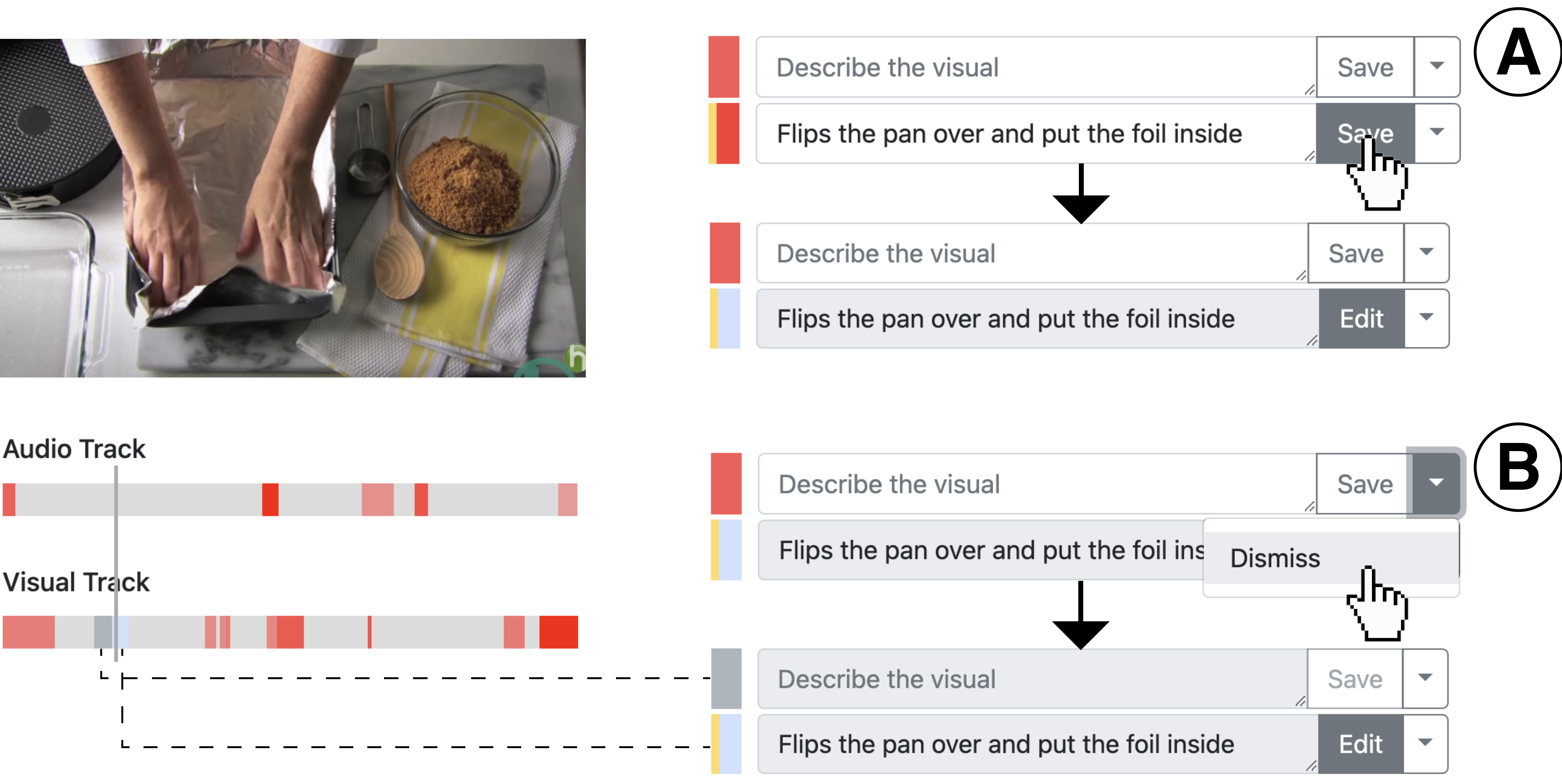}
\caption{(A) After editing their description, authors can add the description to the video by clicking on ``Save''. The vertical side bar and the horizontal bar of the corresponding segment will change to blue color, indicating that this problem has been addressed. (B) Authors can also dismiss a problem by clicking on ``Dismiss''. The bars will turn gray indicating that this problem has been dismissed.}
\label{fig:save_edit}
\end{figure}
The \textit{video description pane} (Figure~\ref{fig:interface}F) lets address inaccessible visual segments by writing text descriptions of the visual content. 
Each video description segment consists of a vertical sidebar that is colored according to its predicted accessibility, an editable text field where an author may add descriptions, and ``save''/``edit'' and ``dismiss'' buttons to address or ignore surfaced visual accessibility issues. 
Video description segments are relatively aligned with the caption segments in the \textit{captions pane} such that authors can preview the nearby narration. The height of each video description segment represents its relative length such that authors can estimate the approximate length of description required. 

When an author locates a visual accessibility issue (\textit{e.g.}, the last displayed segment in Figure~\ref{fig:interface}H), the author can click the segment to play the clip and to check if the visual content is described in the audio or existing descriptions. For example, in this case, the author may notice that the host placing the foil inside the pan is not yet described in the captions, and add a description by typing ``Flip the pan over and put the foil inside'' and clicking ``Save'' (Figure~\ref{fig:save_edit}A). The vertical sidebar, and the corresponding segment in the video pane's visual timeline, then change to blue to indicate the issue has been addressed.
If an author decides that a suggested visual accessibility issue does not need a description, they can dismiss the problem (Figure~\ref{fig:save_edit}B). The vertical sidebar for that segment and the corresponding horizontal bar in the visual track timeline will turn dark gray to indicate the issue has been dismissed.
Authors can manually add a description to a point in the video where an accessibility issue was not detected by double-clicking the visual segment in the video pane's visual timeline to create a corresponding video description segment in the video description pane.

By default, the video description pane displays visual segments with estimated accessibility scores lower than 0.35 (range 0-1). Authors can use the slider (Figure~\ref{fig:interface}G) to surface more visual accessibility problems when making sure they covered everything, or fewer accessibility problems when prioritizing for a time constraint.

\subsection{Captions Pane}
\label{captions_pane}
The \textit{captions pane} lets authors navigate the video with a time-aligned transcript and write captions to address inaccessible audio segments. \systemnamespace automatically provides captions for the speech, so authors can focus on making non-speech sounds accessible. Each caption pane segment has a similar structure with video description segments. Authors can quickly locate, review, script captions in-place, or dismiss a suggested problem. 

The caption pane displays predicted audio accessibility by coloring the vertical bars (similar to the video description pane) to help authors understand and prioritize audio accessibility issues.
Unlike the video description pane, we do not use predicted audio accessibility to filter audio accessibility issues as Closed Caption guidelines state that all important sounds should be synchronously described whether they can be inferred from the visual content alone~\cite{acbguidelines, adlabguidelines}. 
Using the captions pane, authors can click on a caption segment to hear the segment, then script a caption if the sound is important (music at 0:00 in Figure~\ref{fig:interface}E) or dismiss the segment if the sound is not important (silence at 00:25 in Figure~\ref{fig:interface}E).

\subsection{Accessible Video Preview}
After authors create captions and video descriptions for inaccessible segments, they can then preview their results as the video plays. Original captions of the video and captions created by the author are displayed in ``Captions Preview'' (Figure~\ref{fig:interface}D). Audio descriptions are synthesized via a text-to-speech engine (Web API's SpeechSynthesisUtterance Interface\footnote{https://developer.mozilla.org/en-US/docs/Web/API/SpeechSynthesisUtterance}).
Our system renders audio description in the format of \textit{extended description}~\cite{wcagad}, which pauses the video, plays the synthesized speech descriptions, and continues the video.

\section{Cross-modal Grounding Pipeline}
\label{algorithmic_methods}

\begin{figure}[t]
\centering
\includegraphics[width=\linewidth]{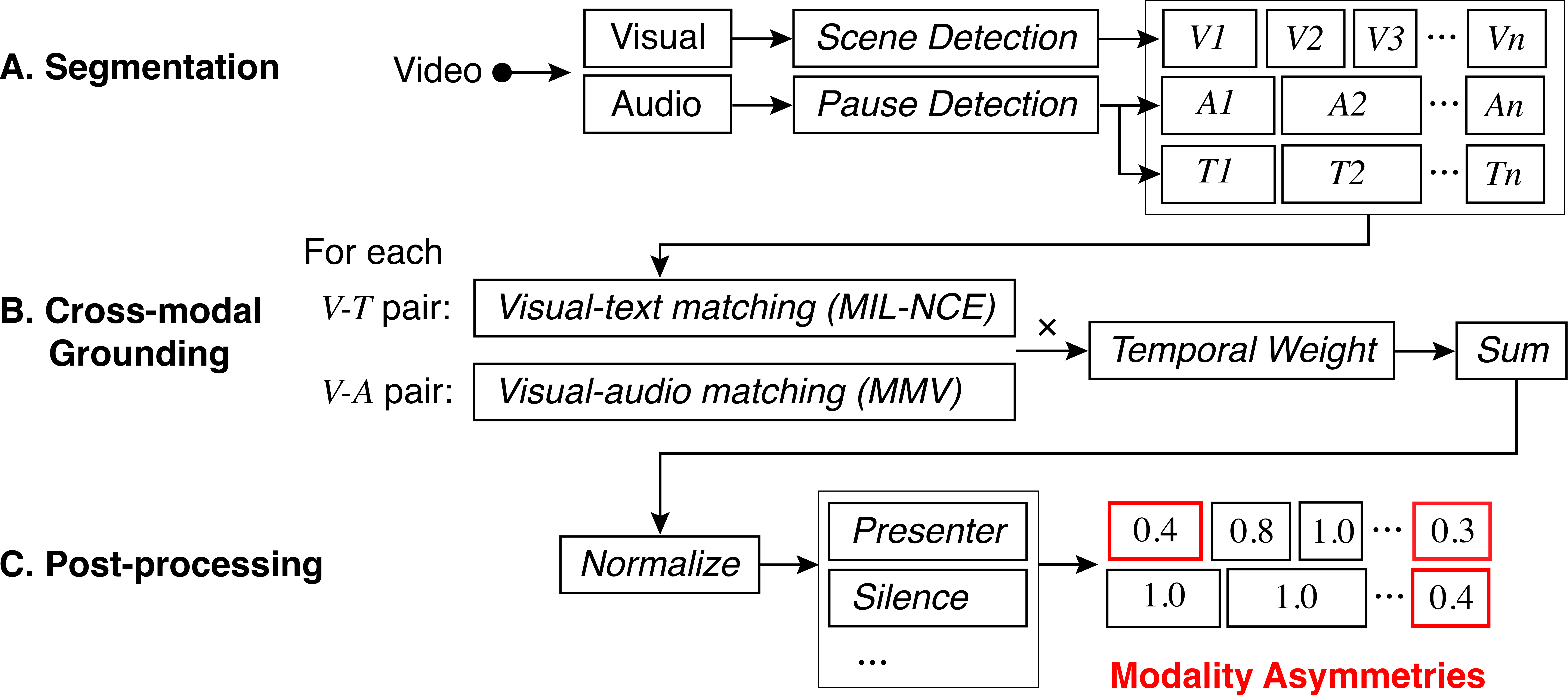}
\caption{Our computational pipeline includes: (A) Segmentation that segments the video into audio and visual segments, (B) Cross-modal Grounding that finds correspondences between visual and audio segments, and (C) post-processing that filters identified correspondences.}
\label{fig:pipeline}
\end{figure}

We present a computational pipeline that segments the auditory and visual track of the video (Figure~\ref{fig:pipeline}A) and identifies asymmetries between auditory and visual tracks using cross-modal grounding analysis (Figure~\ref{fig:pipeline}B). 

\subsection{Segmentation}
To create visual segments, we detect shots, or segments with continuous footage. To segment shots, we used scenedetect\footnote{http://scenedetect.com/}'s content-aware scene segmentation algorithm that compares the HSV color space in adjacent frames against a threshold to determine if the two segments belong to the same shot.
To create audio segments, we follow prior work~\cite{liu2021makes, pavel2020rescribe} by aligning the transcript and audio using Gentle forced-aligner~\cite{ochshorn2017gentle} to get word-level timings, and then consider any gap between words longer than 2 second to be a non-speech audio segment, and any gap longer than 0.5 second but shorter than 2 seconds as a pause in speech. We segment the audio into speech and non-speech audio clips based on the gaps. In addition, we also generate a list of segmented transcripts according to the start/end of audio segments.

\subsection{Cross-modal Grounding}
\begin{figure}[t]
\centering
\includegraphics[width=\linewidth]{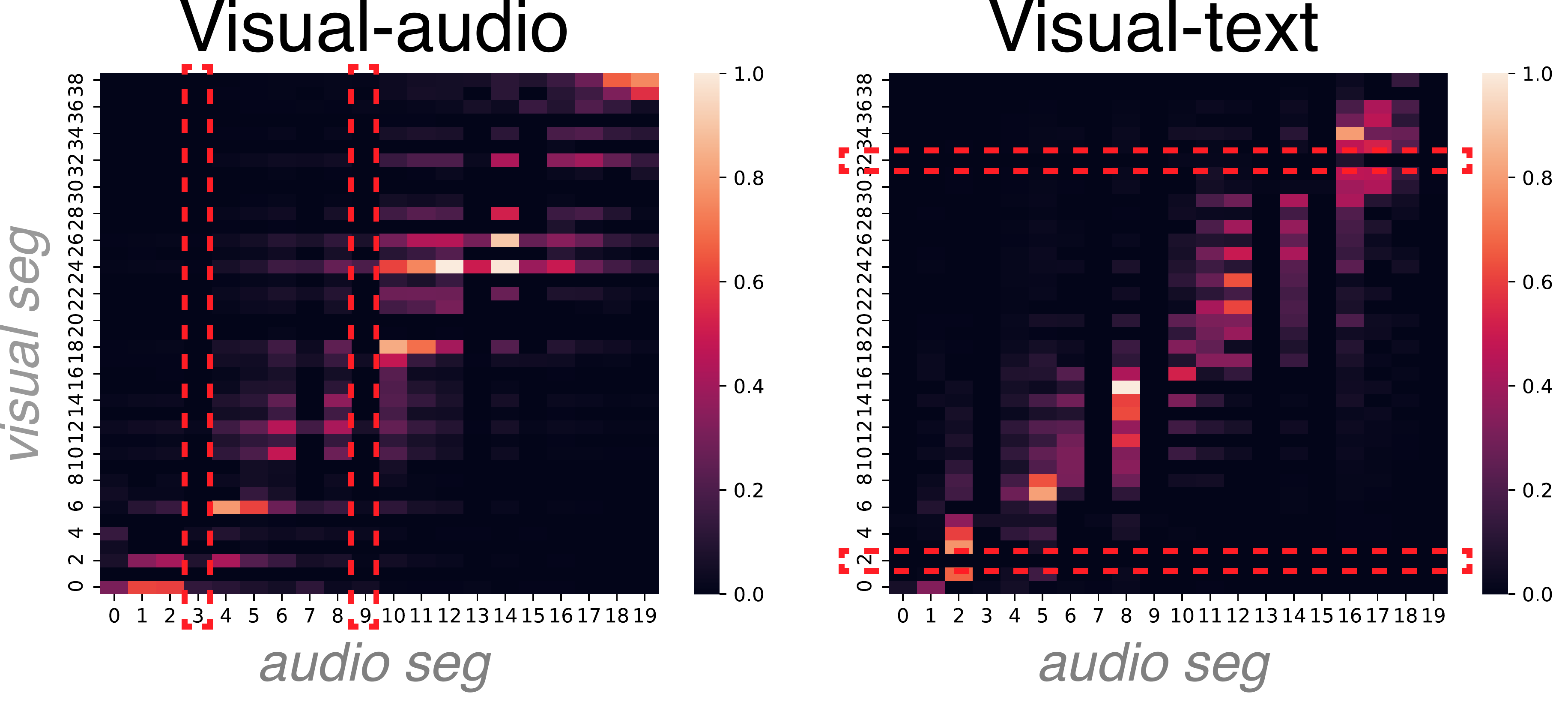}
\caption{Example of visual to audio similarity matrix, and visual to text (transcript) similarity matrix. Red dotted lines highlight examples of asymmetric and potentially inaccessible segments.}
\label{fig:matchings}
\end{figure}
To assess if each visual segment and audio segment is described in the other modality, we compute visual-text and visual-audio matching scores for all video and audio segment pairs using multimodal machine learning algorithms.

\subsubsection{Visual-text/audio Matching}
\systemnamespace uses multimodal machine learning algorithms which learn a symmetric joint embedding space for visual and auditory or textual data. 
With the embeddings, we can measure if a visual segment and a audio segment is semantically similar by computing the dot product of visual embeddings and audio/text embeddings. Specifically, for visual-text matching, we use the MIL-NCE model~\cite{mle2020end} which was trained on HowTo100M, a dataset of 100 million clips-narrations from YouTube. 
For visual-audio matching, we use the state-of-the-art MultiModal Versatile (MMV) networks~\cite{mmv2020self} which was trained on AudioSet, a dataset consists of 10 seconds clips coming from 2 million
different internet videos. AudioSet contains a variety of audio types including musical instruments, animal, mechanical sounds, etc. 

By matching all \(n_v\) visual segments to \(n_a\) audio segments in a video, MIL-NCE and MMV each produces a \(n_v \times n_a\) matrix, where cell \((i, j)\) is the matching score for visual segment \(v_i\) and audio segment \(a_j\). Each matrix is normalized to range \(0-1\). 
Figure~\ref{fig:matchings} displays examples of such cross-modal grounding results. 

To estimate the accessibility of a visual segment, \(v_i\), we compute its matching scores to all audio segments in the same video. When matching to audio segments that contain speech, we use the MIL-NCE score (since match different visuals to human speech sound does not make sense). We remove stop words of all transcripts before computing its correspondence to visuals. When matching to audio segments without speech, we use the MMV score. We then compute a weighted sum of all scores based on each audio segment's temporal position to \(v_i\) (as explained in Section~\ref{temporal_weighting}). Thus, for a video with \(n_v\) many visual segments and \(n_a\) audio segments: 

\begin{equation}
  \operatorname{score}(v_i)=
    \sum_{j}^{n_{a}} w_{i, j} * \operatorname{matching}(v_i, a_j)
\end{equation}

where,

\begin{equation}
  \operatorname{matching}(v_i, a_j)=\begin{cases}
    \operatorname{MIL-NCE}\left(v_{i}, t_{j}\right), & \text{if speech}. \\
    
    \operatorname{MMV}\left(v_{i}, a_{j}\right),    & \text{if non-speech}.
  \end{cases}
\end{equation}

Similarly, to estimate the accessibility score of an audio segment, \(a_j\), we compute its degrees of matching to all visual segments. If the current audio segment is non-speech, we only use the MMV score. 
However, when checking whether the content of a speech audio is presented in the visual, it is prevalent that the speech is transcribed and added as subtitles using automatic speech recognition technology. Even if subtitles of speech are not added, systems can quickly apply ASR to incorporate them into the visual modality. Thus, if an audio segment is detected as speech, we will assign it a constant value \(c\) and consider it accessible (since the speech information is displayed as subtitles in visual). For an audio segment, \(a_j\):

\begin{equation}
  \operatorname{score}(a_j)=\begin{cases}
    c, & \text{if speech} \\
    
    \sum_{i}^{n_{v}} w_{j, i} * \operatorname{MMV}\left(a_{j}, v_{i}\right), & \text{if non-speech}.
  \end{cases}
\end{equation}

Note that because both the MIL-NCE and MMV scores are symmetric (e.g., \(\operatorname{MMV}\left(a_{j}, v_{i}\right) = \operatorname{MMV}\left(v_{i}, a_{j}\right)\)), we only need to perform matrix multiplications one time to compute scores for both visual and audio segments. 

\subsubsection{Temporal Weighting}
\label{temporal_weighting}
A visual segment can be matched to an audio segment that is far away from each other in time. In such cases, although the information is grounded in the other modality, it would be hard for people to connect and make sense of such cross-reference. Thus, as shown in the above equations, we apply a temporal weighting to the output visual-text and visual-audio matching scores. Specifically, the matching between a visual segment \(v_i\) and an audio segment \(a_j\) diminishes exponentially by a factor of \(w\) (\(0\leq w \leq1\)) for every 5 seconds' distance in time: 

\begin{equation}
    w_{i, j}=w_{j, i}=w^{\frac{\left|T S_{i}-T S_{j}\right|}{5}}
\end{equation}

Where \(TS_i\) is the timestamp in seconds of segment \(i\), and \(w\) is the weighting factor. We empirically found that \(w = 0.45\) works well.

\subsection{Post-processing}
\label{postprocessing}
After we compute segment visual accessibility scores, \(\operatorname{score}(v_i)\), and audio accessibility scores, \(\operatorname{score}(a_j)\), for the video, we normalize the scores into \(0-1\) ranges. We then remove commonly detected accessibility issues that do not need further description including: the presenter speaking to the camera, and silences.

\subsubsection{Presenter Speaking}
In initial tests, our approach detected moments where the host was speaking to the camera to be inaccessible, leading to low precision for how-to and recipe videos like \cite{youtubeA} (precision=0.356, recall=0.875) and \cite{youtubeB} (precision=0.435, recall=0.929). While the visual content and speech did not match, these segments were accessible as they could be implied from the audio alone (the presenter's voice).
To address this, we detect faces using OpenCV\footnote{https://docs.opencv.org/3.4/db/d28/tutorial\_cascade\_classifier.html}, and compute the area of the detected face bounding box per second for each visual segment. We consider a visual segment to be ``presenter speaking'' and thus not inaccessible if the area per second metric is greater than a threshold \(TH_{presenter}\). We empirically determined a threshold \(TH_{presenter} = 58000\). 
With presenter detection, the precision improved to 0.636 on \cite{youtubeA} and 0.867 on \cite{youtubeB}.
\subsubsection{Silences}
Similarly, our initial approach predicted segments with silent or insignificant audio to be inaccessible as the quiet noises were not detected to match the visuals.
For instance, the algorithm considered a scene in a recipe video~\cite{youtubeA} where the host flips the pan over making minor noises, and a scene in a food review video~\cite{youtubeC} where the host was eating and making a chewing sound, as inaccessible.
To address the issues, we detect silences by computing 
the average intensity of audio segments using librosa\footnote{librosa.org} and compare it to a threshold \(TH_{silence}\), which we empirically set to 0.007. If the average intensity score is lower than the threshold we label this audio segment as insignificant and thus not inaccessible. 

\subsubsection{Threshold}
As discussed in~\ref{video_description_pane}, \systemnamespace displays visual segments with grounding score larger than a threshold as the visual accessibility issues. We selected 0.35 empirically as it worked consistently well for diverse videos, and favored recall over precision such that the system showed more potential accessibility issues. Authors may easily dismiss false accessibility issues using the ``Dismiss'' button. 

\subsection{Technical Evaluation}
\label{technical_evaluation}
\begin{table}[ht]
\centering
\resizebox{\linewidth}{!}{%
\begin{tabular}{@{}lrrrrrr@{}}
\toprule
 &
  \multicolumn{1}{l}{} &
  \multicolumn{1}{l}{\textit{Visual}} &
  \multicolumn{1}{l}{} &
  \multicolumn{1}{l}{} &
  \multicolumn{1}{l}{\textit{Audio}} &
  \multicolumn{1}{l}{} \\ \midrule
 &
  \multicolumn{1}{l}{Random} &
  \multicolumn{1}{l}{Gaps} &
  \multicolumn{1}{l}{\textbf{\systemname}} &
  \multicolumn{1}{l}{Random} &
  \multicolumn{1}{l}{Gaps} &
  \multicolumn{1}{l}{\textbf{\systemname}} \\ \midrule
Precision & 0.275 & 0.833  & \textbf{0.694} & 0.125 & 0.909 &\textbf{0.983}  \\
Recall    & 0.390 & 0.385 & \textbf{0.984} & 0.381 & 0.843 &\textbf{0.843} \\ 
F1        & 0.323 & 0.526  & \textbf{0.814} & 0.188 & 0.874 &\textbf{0.908}  \\ \bottomrule
\end{tabular}%
}
\caption{\systemname's experimental test results on a sample of 20 manually labeled videos. \systemnamespace generally performs better than random guess and using ``gaps in speech'' as heuristics, for both detecting visual and auditory accessibility issues.}
\label{tab:performance}
\end{table}

We evaluated \systemname's cross-modal grounding pipeline using 20 randomly selected videos from YouDescribe\footnote{https://youdescribe.org/}, a platform where people can request audio descriptions for YouTube videos.
In particular, we limited our random selection to videos that were less than 5 minutes with captions available (implies some narration).
All 20 videos were not tested when we built the system, i.e. out-of-bag, and they covered diverse topics: how-to (5 videos), recipe (4 videos), vlog (3 videos), campus tour (3 videos), documentary (2 videos), educational (2 videos) and review (1 video). 

Two researchers independently identified visual and auditory accessibility issues in the videos based on guidelines~\cite{acbguidelines, fcc, wcag2} (Appendix \ref{appendix_guidelines}). 
We first organized our initial labels in a spreadsheet, and held three one-hour long discussion sessions and went through each accessibility problem one by one.

70.78\% of initially identified issues were the same. Researchers discussed the remaining labels until agreement was reached. In total, the sample included 182 visual accessibility issues and 79 auditory accessibility issues. 
There were two major kinds of disagreements: (1) Missing labels. As a reflection of our motivation, identifying visual and audio accessibility issues for 20 videos was time-consuming and mentally demanding. Most of our disagreements were accessibility issues noticed by one researcher but missed by another. We quickly agreed on those issues. (2) Misinterpreting accessibility guidelines. One of our researchers initially thought that a non-speech audio segment does not need to be described as long as it can be inferred from the visuals. We corrected this mistake during the review of our labels.

We then used \systemnamespace to predict visual and auditory accessibility issues in these videos.
All 20 videos and \systemname's demo of those videos are available online\footnote{\url{https://xybruceliu.github.io/CrossA11y/}}. 
We also predicted accessibility issues using two baselines for comparison: (1) mark each segment as inaccessible with 50\% chance (Random)
, and (2) mark each segment as inaccessible if it did not include speech (Gaps) following prior work~\cite{pavel2020rescribe,natalie2021efficacy,yuksel2020human,wang2021toward}. 
To assess whether a labeled visual or auditory accessibility issue was accurately detected, we compared the start and end times of all segments that were predicted to be inaccessible with the start and end times of manually labeled inaccessible segments. 
We defined a prediction as accurate if there was an \(>50\%\) overlapping manually labeled accessibility problem. 

Using the thresholds we selected in Section~\ref{algorithmic_methods}, \systemnamespace achieved a higher F1 score compared to the baseline methods (Table~\ref{tab:performance}).
For visual accessibility issues, the recall score increased significantly from 0.385 with gaps in speech to 0.984 with \systemname, meaning that \systemnamespace identified visuals with accompanying speech that are still inaccessible to BVI audience members. The precision decreased from 0.833 with gaps in speech to 0.694 with \systemname.
On average, for each video (average length = 3 minutes 11 seconds) we detect 9.1 true visual accessibility issues and 3 of these issues are false positives.
For \systemname, we prefer high recall (the ability to show all issues) over high precision (the ability to show few incorrect issues), as authors may easily review and dismiss inaccurate issues (false positives), but they may struggle to find issues that we do not surface (false negatives).
\systemnamespace identified auditory accessibility problems with higher precision (0.983) compared to gaps in speech (0.909), as \systemnamespace removes false positive issues when the gaps in speech correspond to silence. The recall remains the same as all accessibility issues in our sample occurred during gaps in speech.

\subsection{Limitations}
\label{tech_limitations}

From our technical evaluation, we discuss some limitations of the current implementation of \systemname.  

\subsubsection{Segmentation Limitation}
We noticed that algorithms in some cases failed to segment visual and audio tracks into semantically coherent segments. For visuals, the shot detection algorithm would sometimes segment the same visual with different filming angles into different shots. This leads to lots of repetitive segments that would be annoying for authors to dismiss. In addition, the algorithm sometimes consider a long shot with different pieces of information as one large segment. This is especially common for tutorial videos that only has one shot angle (\textit{e.g.,} an origami tutorial where the camera is always facing the table). 

Similarly, the pause detection algorithm also in some cases produces disproportionally long (\textit{e.g.}, a host speaks very fast with no pause) or short (\textit{e.g.}, a host talks slowly very demonstrating a step in a how-to video) segments. Our algorithm also does not address overlapping sounds like a sound effect that is covered by speech, since audio source separation still cannot produce desirable results and often requires training on specific examples. 

Moreover, visual and audio segment is only a proxy of ``information piece'' that we truly want to extract. In future work will explore methods to address these issues and extract more fundamental units of information pieces.

\subsubsection{Grounding Limitation}
From our observations, current cross-modal grounding algorithms do not work well on visual details that are specific to the current video's context (\textit{e.g.,} in a recipe video the host instructs to mix the batter until it looks like ``this'', the model will label this as matched and cannot detect that the specific state of the batter is not described), and smaller or rarer visuals (\textit{e.g.,} sprinkling salt). Cross-modal machine learning algorithms can also sometimes generate in consistent results due to its unexplainability (giving divergent matching scores to similar visuals that are close to each other).

\subsubsection{Video Production Style}
\systemnamespace works better on videos with relatively dense audio and visual information, and are partially inaccessible. For videos with a monotonous visual (\textit{e.g.,} podcast video stream) or audio track (\textit{e.g.,} only background music), our system will still correctly match visuals and audio, only to show that the entire track is not described. In such scenarios \systemnamespace provides minimal information to authors. 

%
%


%
\section{Evaluations: Can CrossA11y Users Efficiently Identify Video Accessibility Issues?}
\label{eval1}
We evaluated \systemnamespace with 12 participants who have video creation experience to compare creating AD\&CC with and without modality asymmetry visualizations.
We want to investigate: \textit{How does \systemnamespace enable authors to efficiently identify and address visual and auditory accessibility issues in videos?}

\subsection{Materials}
\begin{figure}[t]
\centering
\includegraphics[width=\linewidth]{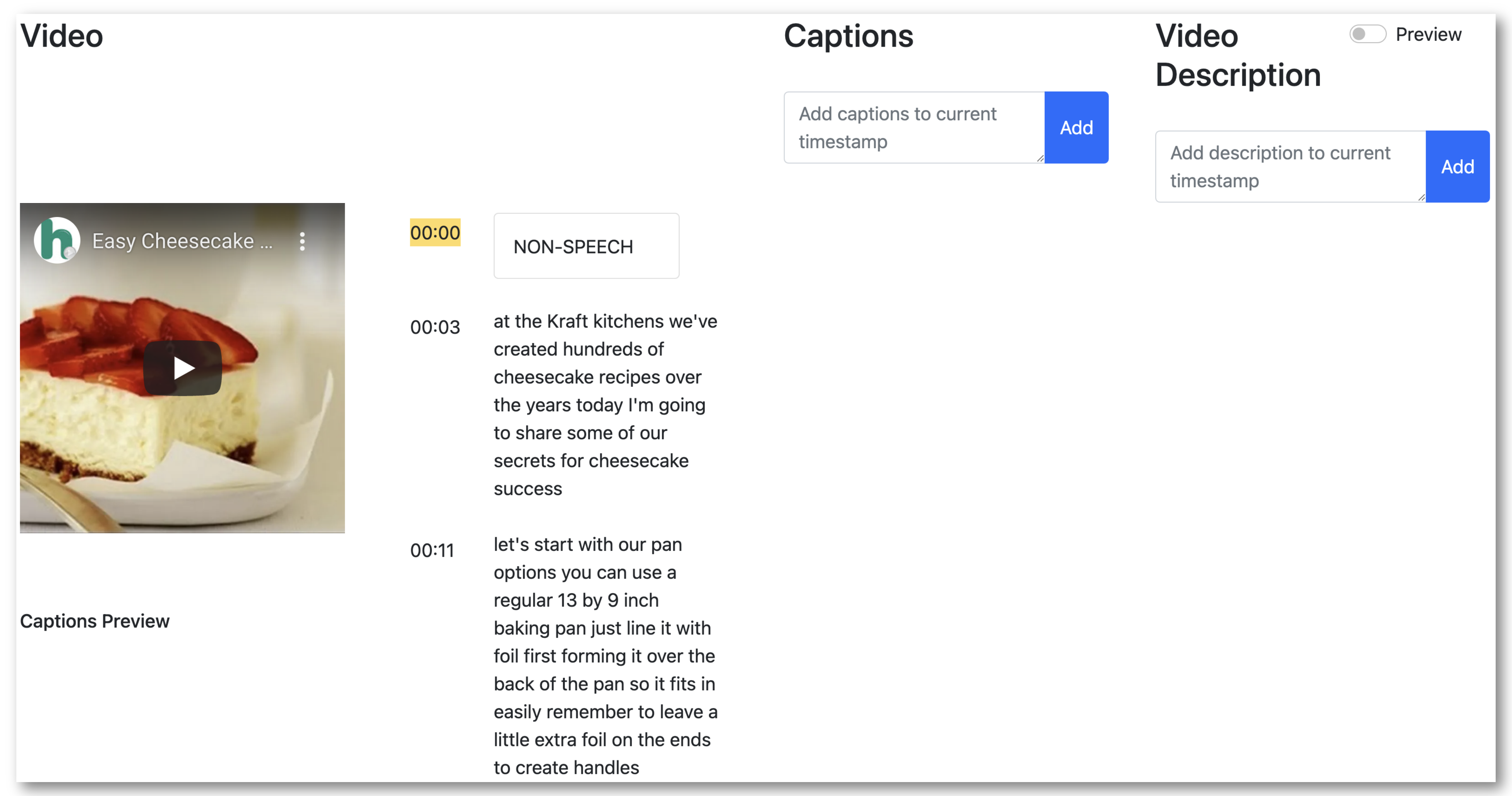}
\caption{\textit{Interface 1} does not provide any accessibility information or modality asymmetry visualization.}
\label{fig:baseline}
\end{figure}

We selected four videos on YouDescribe.com from different genres (Cheesecake recipe~\cite{youtubeA}, handicraft tutorial~\cite{youtubeB}, restaurant review~\cite{youtubeC} and day-in-the-life vlog~\cite{youtubeD}).
All videos are under 5 minutes, and went through the same labeling process as explained in the technical evaluation section.
We used \systemnamespace to automatically identify inaccessible visual and audio segments (Table~\ref{tab:test_videos}) and rendered the four videos on \systemname's interface. 
Additionally, we created \textit{Interface 1} (Figure~\ref{fig:baseline}) where we removed all accessibility visualizations to compare \systemnamespace (Interface 2) with. In Interface 1, we provide a transcript-based timeline and display gaps-in-speech, following prior work~\cite{pavel2020rescribe,natalie2021efficacy,yuksel2020human,wang2021toward}. With Interface 1, users create AD\&CC and click ``Add'' button to add it to the current video timestamp.

\subsection{Participants}
We recruited 12 participants (7 female and 5 male) who all have previous video creation and sharing experience. Participants have created various types of videos including vlog, how-to, music, travel, presentation, product demo, etc. Participants were recruited through our university's internal communication channel and mailing lists.
P8 did not complete the study due to technical issues.
P5, P6, and P11 have their own YouTube/TikTok channels and have created around 40, 80, and 100 videos respectively.
P3, P7, P10 and P12 created 10-20 videos.
P1, P2 P4 and P9 less than 10 videos.

\subsection{Procedure}
We conducted a 90-minute study with each participant remotely. 
Each participant was paid \$50 in gift card.
In each session, we started by asking participants about their experience with videos and experience with accessibility.
We asked if participants have ever added closed captions or audio descriptions to their videos, and their reasons for (not) creating AD\&CC. 
We explained in details what AD\&CC are for, and what they should describe or not describe based on AD\&CC guidelines.
We also reviewed two video examples with AD and CC to provide a more concrete understanding.
Then, we demonstrated Interfaces 1 and 2 with example videos. Each participant tried all features on both interfaces before continuing to the main study. 

Each participant was asked to run four tests in total for Interface 1 and 2 with randomized order.
For each interface, two videos are randomly selected without repetition. 
We provided an open-ended prompt to participants, asking them to \textit{``use this tool to make this video accessible to BVI and DHOH people''}. Participants are not subject to any time limit. 
At the end of each test, we asked participants to rate a set of questions on task load index~\cite{hart1986nasa}. 
After completing all four tasks, we asked participants to compare their experience of identifying and addressing AD and CC with and without \systemname through semi-structured interviews. 

We recorded the audio track for the entire interview and the screen portion of trying out the interfaces.
Both interfaces also automatically logged participants' use of different features, including their video navigation, clicks on vertical and horizontal bars, timestamps they chose to add AD/CC, and the content of AD/CC they wrote, etc.
In total, we collected 4042 log instances of interaction data.
We also recorded the completion time for each task for each participant.  

\subsection{Findings}


\begin{figure}[t]
\centering
\includegraphics[width=\linewidth]{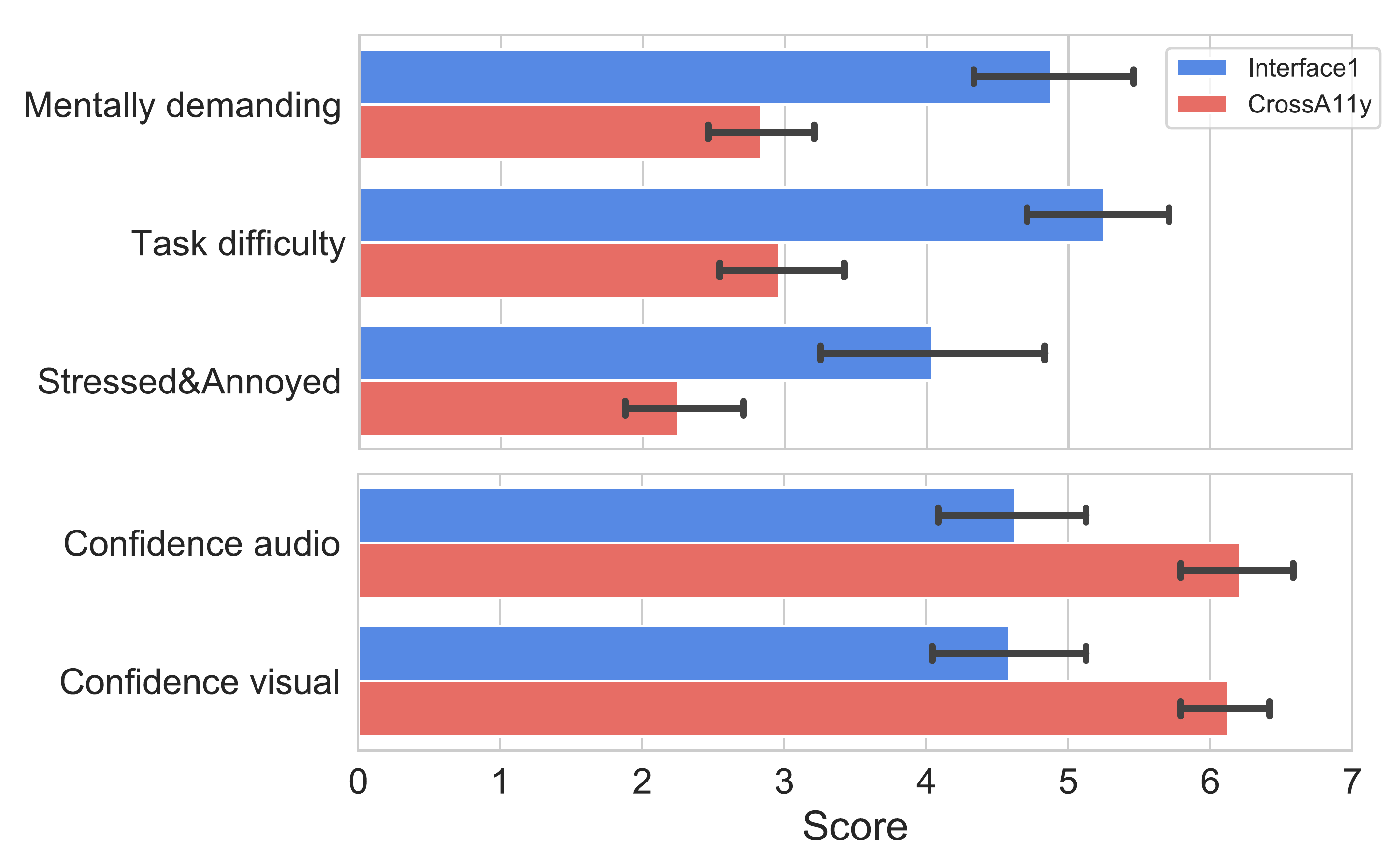}
\caption{Participants' ratings to task load index questions (on a scale of 1-low to 7-high) for their experience adding AD/CC to videos with or without \systemname.}
\label{fig:taskload}
\end{figure}

Participants on average spent 10 minutes and 12 seconds to complete a task.
Participants unanimously preferred using Interface 2 (\systemname) over Interface 1. On a scale from 1-strongly disagree to 7-strongly agree, participants rated that \systemnamespace is useful (\(\mu=6.0, \sigma=0.58\)), easy to use (\(\mu=6.33, \sigma=0.47\)) and would like to use it to make their videos accessible in future (\(\mu=6.50, \sigma=0.65\)). 

\subsubsection{Interface Usage}
With \systemname, most participants navigate the video using the horizontal timelines (Table~\ref{tab:interaction}) in the video pane (Figure~\ref{fig:interface}C) or vertical sidebars in the captions and video description pane (Figure~\ref{fig:interface}E, F). P9 only used the original YouTube player to pause/play the video. She reported that she wanted to be really careful and make sure she covered all problems. 
P1, P4, P6, P7 and P11 used horizontal timelines more frequently to navigate. P4 reported that with horizontal timelines she can more easily understand her current position, and it provided a good overview.
Other participants preferred to navigate in the captions and video description pane. P5 especially liked this ``in-place'' design where he can navigate, identify and edit all in one place. Participants in general navigated more using visual timeline and sidebars than audio. And the frequencies of audio/visual navigation are approximately proportional to the number of audio/visual accessibility issues they addressed. While some participants mostly followed \systemname's guidance, 9 out of 11 participants used ``Dismiss'', ``Add'', and ``Filter'' to correct \systemname's predictions (further discussed in Section~\ref{ai}). 

\begin{table}[ht]
\centering
\resizebox{\linewidth}{!}{%
\begin{tabular}{@{}llrrrrrrrrrrrr@{}}
\toprule
           &                 & P1 & P2 & P3 & P4 & P5 & P6 & P7 & P9 & P10 & P11 & P12 & \textbf{All} \\ \midrule
\textit{Navi} & A timeline  & 14 & 0  & 8  & 16 & 2  & 15 & 12 & 0  & 19  & 13  & 0   & \textbf{99}  \\
           & V timeline & 31 & 0  & 18 & 57 & 3  & 34 & 67 & 0  & 7   & 21  & 4   & \textbf{242} \\
           & A side bar  & 0  & 4  & 15 & 0  & 16 & 0  & 0  & 0  & 16  & 10  & 7   & \textbf{68}  \\
           & V side bar & 0  & 15 & 36 & 1  & 25 & 0  & 10 & 0  & 42  & 14  & 35  & \textbf{178} \\ \midrule
\textit{Edit}    & CC              & 14 & 9  & 12 & 7  & 14 & 10 & 14 & 11 & 12  & 4   & 11  & \textbf{118} \\
           & AD              & 15 & 24 & 26 & 32 & 23 & 25 & 30 & 25 & 36  & 18  & 18  & \textbf{272} \\
           & Edit            & 0  & 1  & 7  & 5  & 2  & 0  & 5  & 2  & 9   & 2   & 2   & \textbf{35}  \\
           & Dismiss         & 7  & 7  & 6  & 11 & 0  & 6  & 7  & 0  & 7   & 2   & 20  & \textbf{73}  \\
           & Add             & 0  & 0  & 3  & 8  & 0  & 0  & 5  & 0  & 2   & 8   & 0   & \textbf{26}  \\
           & Filter          & 0  & 0  & 1  & 0  & 1  & 0  & 0  & 0  & 7   & 0   & 6   & \textbf{15}  \\ \bottomrule
\end{tabular}%
}
\caption{Participants' usage of \systemname's interface for navigation (\textit{Navi}) and addressing video accessibility problems (\textit{Edit}). Participants navigate the video using different approaches, including audio and visual timelines in the video pane (A/V timeline), and side bars of surfaced audio and visual accessibility problems in the captions and video description pane (A/V side bar). Participants in total created 118 captions and 272 audio descriptions in our study.}
\label{tab:interaction}
\end{table}

P1, P5, P8 and P10 particularly liked the use of color in \systemname. Colors allow them to get an overview of approximately how inaccessible this video is, quickly locate the most critical accessibility issues, and monitor their work progress. P8 described:
\begin{quote}
    \textit{``I like that when you have something undone, it will mark as red. This makes it really easy for me to locate the tracks and navigate. You can take a glance at how much work is left. ''} -- P8
\end{quote}

P10 also considered the segmentation of visual and audio to be especially useful. It allows him to add AD/CC to where the scene or sound occurs, and adds it for a coherent piece of information, comparing to Interface 1 that he needed to adjust himself.

\subsubsection{Efficiency and Performance}

Participants on average spent 10 minutes 12 seconds to complete a task.
There is no significant difference between task completion times using Interface 1 and 2 (Table~\ref{tab:participants}).
This could be caused by identifying more accessibility issues and thus spending more time overall to address the issues. For example P5 and P10 reported that Interface 1 required too many efforts and discouraged them from carefully inspecting the videos. 

Thus, instead of comparing the overall time spent, which could be affected by video length, number of accessibility issues identified and efforts to write descriptions, we measure time per fix (Table~\ref{tab:participants}), i.e. total time divided by number of total AD/CC added. This metric represents how efficiently a participant can locate accessibility problems. 
A Wilcoxon signed-rank test shows that participants were able to create AD\&CC more efficiently with \systemnamespace (\(\mu=38.5, \sigma=19.5\)) than with Interface 1 (\(\mu=45.0, \sigma=18.1\)), 
with statistical significance (\(W=182.0, p=0.037\)). Participants found a variety of features \systemnamespace provided to be helpful in making their workflow faster. This will be discussed in details in Section~\ref{identify}

\begin{table}[ht]
\centering
\resizebox{\linewidth}{!}{%
\begin{tabular}{@{}llrrrrlr@{}}
\toprule
 &
   &
  \multicolumn{1}{l}{\begin{tabular}[c]{@{}l@{}}Visual \\ Precision\end{tabular}} &
  \multicolumn{1}{l}{\begin{tabular}[c]{@{}l@{}}Visual \\ Recall\end{tabular}} &
  \multicolumn{1}{l}{\begin{tabular}[c]{@{}l@{}}Audio\\ Precision\end{tabular}} &
  \multicolumn{1}{l}{\begin{tabular}[c]{@{}l@{}}Audio \\ Recall\end{tabular}} &
  Time &
  \multicolumn{1}{l}{Sec/fix} \\ \midrule
\textit{Interface 1}        & mean & 0.631          & 0.621          & 0.589          & 0.487          & 21:18          & 45.0          \\
                            & std  & 0.105          & 0.279          & 0.251          & 0.301          & 4:15           & 18.1          \\
\textit{\textbf{CrossA11y}} & mean & \textbf{0.912} & \textbf{0.895} & \textbf{0.766} & \textbf{0.693} & \textbf{19:33} & \textbf{38.5} \\
                            & std  & \textbf{0.091} & \textbf{0.131} & \textbf{0.192} & \textbf{0.196} & \textbf{4:38}  & \textbf{19.5} \\ \bottomrule
\end{tabular}%
}
\caption{Participants' performance (precision and recall of identifying video and audio accessibility problems), and task time (sec/fix, total amount of time fixing a video divided by the number of accessibility fixes) in user study.}
\label{tab:participants}
\end{table}
Another important measurement for efficiency is how well authors were able to identify visual and auditory accessibility issues, specifically, how many of the captions and descriptions they added address an actual accessibility issue (precision), and how many of the total accessibility issues were addressed (recall). We collected participants' log data and computed the precision and recall scores to locate visual and auditory problems using both interfaces (Figure~\ref{tab:participants}). 
We label a participant's created AD/CC to be correct if its added timestamp lies within the start and end timestamp of a manually labeled accessibility issue. 


Participants reached higher precision and recall to identify inaccessible audio (precision: \(\mu=0.766, \sigma=0.192\), recall: \(\mu=0.693, \sigma=0.196\)) and visual segments (precision: \(\mu=0.921, \sigma=0.091\), recall: \(\mu=0.895, \sigma=0.131\)) with \systemnamespace than with Interface 1 (Table~\ref{tab:participants}). Wilcoxon signed-rank tests indicate statistical significance for the increase in participants' performance for both auditory (precision: \(W=123.0, p=0.004\), recall: \(W=120.0, p=0.004\)) and visual (precision: \(W=0.0, p<0.001\), recall: \(W=24.0, p<0.001\)) accessibility problems. This result aligns with participants responses to task load index questions. Participants felt significantly more confident in locating auditory (\(W=35.0, p=0.003\)) and visual (\(W=4.5, p<0.001\)) accessibility issues with \systemnamespace than without (Figure~\ref{fig:taskload}).
5 participants 
reported that \systemnamespace not only provides them with some guidance, but also serves as a confirmation that improves their confidence.

\subsubsection{\systemnamespace Workflows}
\label{identify}
Participants were able to identify and address accessibility problems with \systemnamespace more efficiently. 
Results from task load index questions (Figure~\ref{fig:taskload}) also show that the participants found that using \systemnamespace was significantly less mentally demanding (\(W=9.5, p<0.001\)), less difficult (\(W=10.5, p<0.001\)), and less stressed or annoyed (\(W=11.5, p<0.001\)) than using Interface 1. 

All participants reported that in Interface 1 they had to watch the entire video through and have to constantly check if there is an accessibility issue. P6 complained that she had to \textit{``stop at every sentence''}. P9 had to \textit{``pay attention all the time, every moment''}.
Moreover, 8 out of 11 participants stated that it was a huge cognitive load for them to surface for visual and audio accessibility problems at the same time, because they have to repeatedly switch their minds between imagining ``I cannot see the content'' vs. ``I cannot hear the content''. 

As a result, participants either had to play the video two times and only focus on addressing one type of accessibility issue at a time (P3, P6, P7, P9, P11), repeatedly inspect the same segment (P4), or rely on heuristics to make the process easier (P1, P2, P5):

\begin{quote}
    \textit{[With Interface 1] I first look for non-speech, or visual changes to some obvious object or some close-up shots. Then I would imagine that I can only access the video through one of my senses, for example my vision or my hearing, to determine that, ok, here might need a CC or AD. } -- P2
\end{quote}

\systemnamespace enabled participants to more efficiently locate and address accessibility problems with lighter mental demand. With visualization of modality asymmetries, participants can get an overview of, immediately identify, and seamlessly navigate to surfaced visual and auditory accessibility issues.  
8 out of 11 participants would directly jump to the highlighted red visual and audio segments in a video and address those problems, especially after they felt that the algorithm is accurate enough:

\begin{quote}
    \textit{``After the first one I felt like the algorithm is pretty accurate and sufficient. So in the second one, I would just click on the undone red marks. It’s a much better experience.''} -- P3
\end{quote}

While \systemnamespace highlights inaccessible segments so that authors can directly jump to these segments and write descriptions, one limitation of this workflow is that authors may sometimes miss important context of a video. No participants in our study reported difficulty writing descriptions due to lack of context. We analyzed a sample of descriptions created by participants during the study, no major completeness or accuracy issue was found. This could be due to the selected testing videos that do not heavily rely on context (tutorial, recipe, vlog, review), compared to other forms of content like stories or lectures.

Interestingly, P2, P6 and P10 employed a ``dynamic workflow'' in which they would still skim through the gray segments while paying more attention to the red parts: 
\begin{quote}
    \textit{``My workflow isn’t linear anymore and I don’t have to check for every second. For example, in this video I can click on a gray segment to instantly navigate to the position, and then realize that most of it is just the person speaking, then I can just skip the entire segment and go to the next one. For the first one I’ll have to be continuously watching.''} -- P6
\end{quote}

P2, P7 and P10 also explained that \systemname's highlight of inaccessible segments reduces the work from searching for all potential accessibility issues to judging if one modality of this video segment is inaccessible, which is much less mentally demanding:
\begin{quote}
    \textit{``Seeing the problem, I can understand it in hindsight. It is so much easier than I have to go over everything, paying attention to visual and audio, thinking if there is potentially an accessibility issue while the video is still playing.''} -- P7
\end{quote}

9 out of 11 participants reported that with \systemnamespace they were able to address both visual and auditory problems in parallel. P4 described that with the timelines she realized that most audio and visual issues are not in the same location. So when she was at a segment she can focus on either audio or visual. And even if they are around the same location, P4 explained, \textit{``Since you know that there might be a problem, your attention will be on what potential problem does this segment have instead of which part has a problem. I don't have to distinguish. I feel like that was the hardest part.''}

\subsubsection{Interpreting AI Predictions}
\label{ai}
9 out of 11 participants used ``Dismiss'', ``Add'', and ``Filter'' to correct \systemname's predictions. As discussed in previous sections, participants can easily judge and dismiss false positive predictions by our system.  
We observed that participants were able to identify visual accessibility issues with significantly higher precision (0.921), compared to the precision of \systemname's predictions (0.718). This indicates that participants were not overly relying on the system and were able to determine whether the surfaced problem is actually inaccessible. P2, P5, P6, P9 and P10 also checked for false negative errors of \systemname, by skimming through gray segments or adjusting the slider (Figure~\ref{fig:interface}H) to retrieve more accessibility problems:
\begin{quote}
    \textit{``After I have address all the issues, usually I just slide it to a bigger value and check if there’s any red segments. If there is I’ll click on those segments and see if they are actually accessible. If all the new problems are ok I’ll stop there.''} -- P10
\end{quote}

P1, P2 and P5 reported that they would prioritize workload over \textit{complete} accessibility. P2 stated that she will first look at the top-left corner to see how many issues remaining and just go with the default if not too many or not too few. P5 told us that having some description to cover some important visual stuff in his video is more important than completeness:

\begin{quote}
    \textit{``Using this tool, I’m not trying to achieve a 100\% accuracy. For any suggestions it provides, it’s already better. it increases my willingness to address them and at least try to make my videos more accessible. If I’m using the first one [Interface 1] I’ll probably just choose to skip.''} -- P5
\end{quote}

\subsubsection{Feedback \& Improvement}
During the study, participants suggested new features that could improve our interface. P1 and P4 thought that sometimes segmented visual and audio content is repetitive and they have to enter the same descriptions again and again. They suggested that we could cluster similar visuals/audio segments and allow authors to apply a description to all similar segments. 

Although our system focuses on the identification of accessibility issues, as a number of prior work~\cite{saray2011adaptive, natalie2021efficacy, peng2021say, pavel2020rescribe} have explored ways to author higher quality descriptions, 7 out of 11 participants hope that our system could automatically generate descriptions and captions, or at least some simple words to start with. They felt like this is the ``last missing piece'' of our system and would consider use it on all of their video creations.

P10 felt that our current design of the slider is a bit hard to understand, and we could potentially replace it with more well-defined levels. For instance, ``You should fix this'', ``Recommend fixing'', ``Make your video completely accessible'', etc. 
 \section{Case Study: Usage of \systemnamespace by Content Creators}

We recruited two YouTubers who did not participate in the first study and conducted a 60-minute study with each participant remotely.
Each author was paid \$50 in gift card. 
We demonstrated \systemnamespace and asked the authors to make two of their own videos\footnote{
\url{https://www.youtube.com/watch?v=TMxZ7vooRm8},\\ \url{https://www.youtube.com/watch?v=r3v7GFBhWdI}\\
\url{https://www.youtube.com/watch?v=lS35Yq_dv9k}\\
\url{https://www.youtube.com/watch?v=fRoegT3_Z_E}} accessible using our system. We then conducted a semi-structured interview to discuss their experience, concerns, and expectations. Specifically, we wanted to understand: \textit{How could \systemnamespace fit within content creators’ video creation workflow, and help them make their own videos more accessible in the real-world settings?}


\subsection{Participants}
Author 1 mainly created life vlogs and music videos, with around 70 videos published on YouTube. Author 2 mainly created life vlogs and talking videos, with around 200 videos published in YouTube and TikTok. Only the second author has added closed captions to her videos before (not often). Neither of them has added audio descriptions to their videos.
\subsection{Findings}

\subsubsection{Integrating \systemnamespace into Video Production}
Both authors gave a 7 when rated on the usefulness of \systemnamespace (from 1-strongly disagree to 7-strongly agree). The authors agreed that \systemnamespace helps identify inaccessible parts while doing it manually, even in their own videos, is difficult for them.
Both authors expressed enthusiasm and showed strong expectations towards integrating \systemnamespace into existing video uploading process. Author 1 mentioned that she might consider how she designs the production style (\textit{i.e.,} more description of the visual) of her video content based on accessibility feedback: 
\begin{quote}
    \textit{``If it's integrated within my editor, I would try to edit the video in a way that's more accessible. It would help me keep an eye out for parts that are particularly inaccessible, or if I noticed that a lot of scenes are inaccessible, I might rethink how to structure the video better.''} -- Author 1
\end{quote}
However, author 2 would still prioritize what she wants to express first when producing the videos, and would only consider accessibility until the editing is complete. She preferred to have a completely separate tool or website that she could upload the video and check for issues before she published it onto YouTube.

\subsubsection{Balancing Efforts and Accessibility} Both authors gave a 6 when rated on `` I would like to use \systemnamespace to check and address for accessibility issues in the future.'' (from 1-strongly disagree to 7-strongly agree) Their major concern that stops them from using this tool is balancing between time invested in fixing accessibility issues and how many audiences indeed benefit from it. 
Author 1 reported that she did no know how many BVI or DHOH people were watching her videos and required AD/CC, so she had not thought about making her videos more accessible. However, if she knew that even a small part of her demographic needed it, she would definitely invest some time and use this tool to make her content accessible in the future. 
Author 2 explained that she would try to make her video as accessible as possible with the time she had:

\begin{quote}
    \textit{``I always check my video through a color accessibility test, because it's something that takes a short amount of time, but allows my videos to be a little more accessible. So if your tool were to come out I will definitely do it because it's gonna take me less than 30 minutes for two videos while 45 minutes for one video manually [adding closed captions]. It saves me so much time.''} -- Author 2
\end{quote}

\section{Discussion and Future Work}
Our work explores using cross-modal grounding to detect modality asymmetries in the visual and auditory tracks in a video, and instantiates the scores as a unified interface that allows users to efficiently identify and address video accessibility issues. 
Next, we describe the limitations of our system, discuss the implications, and envision future opportunities: 


\textbf{Improved Segmentation of Information. }
In \systemnamespace, visual and audio tracks of media are first divided into semantically coherent pieces of information.
In our implementation, we chose to use pauses in speech to segment the audio track and shot changes to segment the visual track. However, this proxy can be inaccurate sometimes. 
In the future, we hope to explore more semantically meaningful 
approaches to segment visual and audio information. For example, we will segment visuals into object-level segments
such that each segment corresponds to one important visual object.

\textbf{Leveraging Information Importance.}
In our current implementation, \systemnamespace only detects unmatched segments and does not discern if an unmatched visual/audio segment contains important information. This results in the presenter (\eg a host talking to the camera does not contain much useful information) and silence issues (\eg silence or background noise does not to be explicitly described) that we have to address.
Future systems could estimate how important an unmatched visual/audio is (\textit{e.g.}, based on the topic's uniqueness or consistency with respect to the rest of the video~\cite{pavel2020rescribe}), then surface and prioritize segments for authors based on importance in addition to accessibility. 

\textbf{Incorporating \systemnamespace with Existing Systems.}
7 of 11 participants in our lab study and both content creators mentioned that they wanted automated AD/CC as a starting point. Although this research is focused on helping users identify accessibility problems rather than authoring AD/CC, we see an opportunity to combine \systemnamespace with existing systems like ~\cite{saray2011adaptive, natalie2021efficacy, peng2021say, pavel2020rescribe} to create an end-to-end experience for authors. 

However, researchers should also be cautious when providing AI-generated info as replacement of content authoring since it may decrease the content quality.
Prior research \cite{10.1145/3441852.3471207} on alt text authoring showed that authors wrote significantly lower quality alt text when starting with automatic alt text compared to starting with a blank box. We see one major opportunity for researchers to design the representation of AI-generated info to assist content authoring without priming authors with low quality automatic generations.

\textbf{Modality Asymmetry for Accessibility Beyond Video.}
We use cross-modal grounding to check for modality asymmetries in visual and audio. 
In the future, we will apply our cross-modal grounding pipeline to other forms of media. As long as the context of the media includes two or more modality sources (images and its article, gifs and its post text), we can use the modality asymmetry to identify the inaccessible parts. For example, the pipeline can segment images using semantic segmentation, object recognition and vision-language models, and then detect which part of an image is not described by the text description, presenting accessibility issues to BVI people; or which part of the text description is not represented in the image, presenting accessibility issues to people with Dyslexia or people who do not understand the language.

Future research can explore generalizing this pipeline. 
Since most media consists of information from three main modalities, visual, auditory and textual, there is an opportunity to design a unified system that is able to provide an accessibility diagnosis for all major media contents with a consistent standard.


\textbf{Empathize and Incentivize Accessibility.}
The majority of participants from both evaluations claimed that the small number of BVI/DHOH audience, the lack of system assistance and video platform support are the reasons why they did not provide AD/CC for their created videos.
Without assistance and support, they had to invest a lot of effort when they were not sure if someone in their audience could benefit from the effort.
In fact, on YouDescribe\footnote{https://youdescribe.org/}, a crowdsourcing audio description platform, blind and low vision people submit a large amount of requests every day to make YouTube videos accessible. 
Providing more specific instructions or reminders on the video platform can help incentivize authors to add CC/AD. As Author 2 commented, \textit{``Your system should educate authors about what is accessibility and why it's important.''} 

\section{Conclusion}

We present \systemname, a system that enables authors to efficiently identify and address visual and auditory accessibility issues in videos. Our system automatically estimates accessibility of visual and audio segments by checking for modality asymmetries using cross-modal grounding algorithms. It allows authors to quickly locate, review, script and preview AD\&CC in a unified interface. Participants using \systemnamespace in our user studies were able to author AD\&CC more efficiently with lower mental demand. Content creators envisioned integrating our system into their video creation workflow, and expressed enthusiasm in using it to make their videos more accessible in the future.

\bibliographystyle{ACM-Reference-Format}
\bibliography{refs}

\appendix

\begin{figure}[t]
\centering
\includegraphics[width=0.8\linewidth]{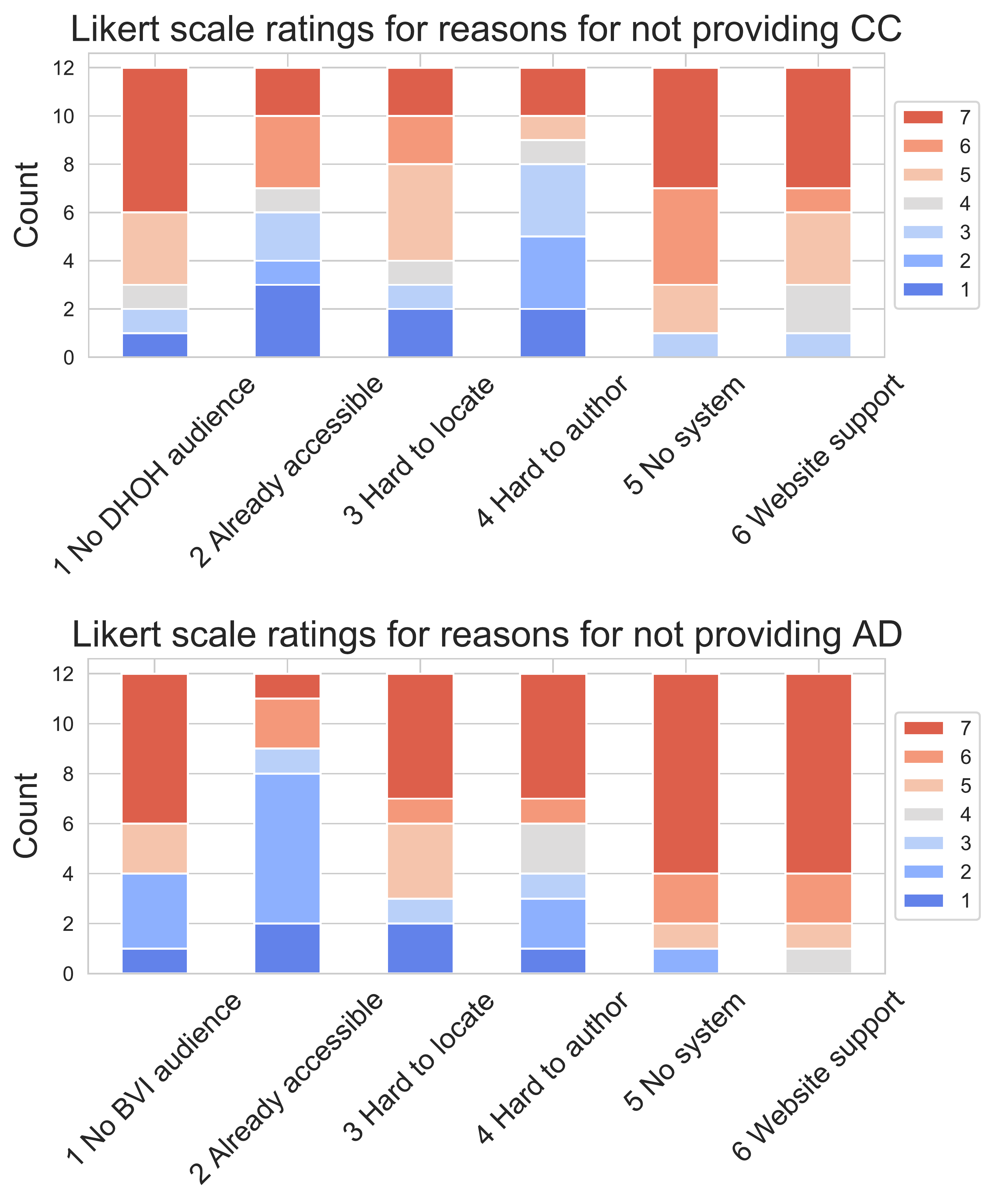}
\caption{Participants' responses to ``I did not provide AD/CC for my created videos because: 1) I don't think there's BVI/DHOH people watching my video. 2) My video is already visually/auditorily accessible. 3) I find it hard to locate visual/auditory accessibility issues. 4) I find it hard to write AD/CC. 5) There is no system that can assist me in creating AD/CC. 6) Video platforms' support is not good.''}
\label{fig:why_no_cc_ad}
\end{figure}

\section{Summary of guidelines we used to manually identify visual and auditory accessibility issues}
\label{appendix_guidelines}

We provide a summary of the guidelines we used to manually identify visual and auditory accessibility issues.
For visual accessibility issues:
\begin{itemize}
    \item Describe visual content that is important to understand the video including objects (e.g., an ingredient for a recipe), actions (e.g., a recipe step), and scenes (e.g., the kitchen).
    \item Do not describe details that are understandable from the audio alone (e.g., the camera returns to a host speaking to the camera).
\end{itemize}

For audio accessibility issues:
\begin{itemize}
    \item Transcribe speech (in our case, it was already available from the existing captions).
    \item Describe non-speech sounds including: relevant environmental sounds (e.g., blender, alarm clock), background music, and sound effects (e.g., a rimshot to accent a joke)
\end{itemize}

\section{Evaluation 1}
\subsection{Detailed Participants' Performance Table}
\begin{table*}[ht]
\centering
\resizebox{\textwidth}{!}{%
\begin{tabular}{@{}llrrrrrrrrrrrr@{}} \toprule
     &                & \textit{Interface 1} &       &       &       &       &      & \textit{\textbf{CrossA11y}} &       &       &       &       &      \\ \cmidrule{3-8} \cmidrule{9-14}
P\# &
  Tasks &
  V Precision &
  V Recall &
  A Precision &
  A Recall &
  Task Time &
  Sec/fix &
  V Precision &
  V Recall &
  A Precision &
  A Recall &
  Task Time &
  Sec/fix \\ \midrule
P1   & B1, D1, A2, C2 & 0.786    & 0.407 & 1.000 & 1.000 & 14:30 & 48.3 & 0.929     & 0.765 & 0.583 & 0.875 & 11:03 & 26.5 \\
P2   & A1, D1, C2, B2 & 0.786    & 0.239 & 0.778 & 0.636 & 21:58 & 69.4 & 1.000     & 1.000 & 0.555 & 0.500 & 23:34 & 45.6 \\
P3   & C1, B1, D2, A2 & 0.636    & 0.304 & 0.375 & 0.300 & 13:33 & 42.8 & 0.850     & 0.810 & 0.909 & 0.909 & 08:20 & 16.7 \\
P4   & A1, C1, B2, D2 & 0.611    & 0.647 & 0.875 & 0.875 & 30:47 & 71.0 & 0.964     & 1.000 & 1.000 & 0.462 & 34:22 & 62.5 \\
P5   & B1, D1, C2, A2 & 0.552    & 0.593 & 0.444 & 0.308 & 13:29 & 21.9 & 0.905     & 1.000 & 0.583 & 0.875 & 12:25 & 23.3 \\
P6   & C1, A1, D2, B2 & 0.468    & 1.000 & 0.500 & 0.875 & 31:56 & 31.4 & 0.960     & 0.889 & 1.000 & 0.769 & 12:25 & 21.3 \\
P7   & A1, B1, C2, D2 & 0.694    & 1.000 & 0.333 & 0.250 & 19:40 & 28.1 & 0.889     & 1.000 & 0.636 & 0.778 & 12:51 & 20.8 \\
P9   & D1, A1, B2, C2 & 0.526    & 0.476 & 0.667 & 0.182 & 12:35 & 36.0 & 1.000     & 1.000 & 0.636 & 0.700 & 17:17 & 30.5 \\
P10  & B1, C1, A2, D2 & 0.655    & 0.826 & 0.273 & 0.300 & 21:43 & 38.3 & 0.704     & 0.905 & 0.900 & 0.818 & 31:33 & 52.6 \\
P11  & C1, A1, B2, D2 & 0.538    & 0.412 & 0.400 & 0.250 & 21:52 & 72.9 & 1.000     & 0.593 & 1.000 & 0.308 & 23:55 & 75.5 \\
P12  & D1, B1, C2, A2 & 0.694    & 0.926 & 0.833 & 0.385 & 23:26 & 35.2 & 0.833     & 0.882 & 0.625 & 0.625 & 20:15 & 48.6     \\ \midrule
\textbf{mean} &                & 0.631   & 0.621 & 0.589 & 0.487 & 21:18 & 45.0 & \textbf{0.921}     & \textbf{0.895} & \textbf{0.766} & \textbf{0.693} & \textbf{19:33} & \textbf{38.5} \\
\textbf{std}  &                & 0.105    & 0.279 & 0.251 & 0.315 & 4:15  & 18.1 & \textbf{0.091}     & \textbf{0.131} & \textbf{0.192} & \textbf{0.196} & \textbf{4:38}  & \textbf{19.5} \\ \bottomrule
\end{tabular}%
}
\caption{Tasks performed by each participant (A1 represents that the participant used Interface 1 with video A in Table~\ref{tab:test_videos}), their performance (precision and recall of identifying video and audio accessibility problems), and task time (total amount of time fixing a video divided by the number of accessibility fixes). Orders of tasks are rearranged for table clarity.}
\label{tab:participants_big}
\end{table*}
Table~\ref{tab:participants_big} shows detailed task performance information for all participants in evaluation 1.

\subsection{Participants' Prior Experience with Accessibility}
All participants in our study did not have experience adding closed captions (to describe non-speech sounds) or audio descriptions to their videos. In our pre-study interviews, we asked participants to rate their agreement (from 1-strongly disagree to 7-strongly agree) with reasons that they did not provide AD/CC for their videos (Figure~\ref{fig:why_no_cc_ad}). Participants listed ``no convenient system'', ``platform support'', ``no DHOH/BVI audience'' as their top reasons. In addition, participants reported that locating accessibility issues is harder for them than authoring descriptions, especially for auditory accessibility problems.

\subsection{Test Videos Used in Evaluation 1}
\begin{table}[ht]
\centering
\resizebox{\linewidth}{!}{%
\begin{tabular}{@{}lrrrrrr@{}}
\toprule
 &
  \multicolumn{1}{l}{} &
  \multicolumn{1}{l}{\textit{Visual}} &
  \multicolumn{1}{l}{} &
  \multicolumn{1}{l}{} &
  \multicolumn{1}{l}{\textit{Audio}} &
  \multicolumn{1}{l}{} \\ \midrule
 &
  \multicolumn{1}{l}{Precision} &
  \multicolumn{1}{l}{Recall} &
  \multicolumn{1}{l}{F1} &
  \multicolumn{1}{l}{Precision} &
  \multicolumn{1}{l}{Recall} &
  \multicolumn{1}{l}{F1} \\ \midrule
A. Recipe~\cite{youtubeA}   & 0.636 & 0.875 & 0.737 & 1.000 & 1.000 & 1.000 \\
B. Tutorial~\cite{youtubeB} & 0.867 & 0.929 & 0.897 & 1.000 & 1.000 & 1.000 \\
C. Review~\cite{youtubeC}   & 0.800 & 0.889 & 0.842 & 0.500 & 0.667 & 0.517 \\
D. Vlog~\cite{youtubeD}     & 0.929 & 1.000 & 0.963 & 1.000 & 0.833 & 0.909 \\ \midrule
\textbf{All} &
  \textbf{0.718} &
  \textbf{0.955} &
  \textbf{0.820} &
  \textbf{0.950} &
  \textbf{0.905} &
  \textbf{0.927} \\ \bottomrule
\end{tabular}%
}
\caption{Test videos used in Evaluation 1 and \systemname's performance on these videos.}
\label{tab:test_videos}
\end{table}
Table~\ref{tab:test_videos} shows CrossA11y's performance on selected test videos in evaluation 1.

\subsection{Task Load Index Questions}

We list the set of task load index questions used in Section~\ref{eval1}.
After each task, we ask (from 1-very low, to 7-very high):
\begin{enumerate}
    \item How mentally demanding was the task? 
    \item How hard did you have to work to accomplish your level of performance? 
    \item How insecure, discouraged, irritated, stressed and annoyed were you? 
    
    \item How successful and confident were you in identifying audio accessibility problems?
    \item How successful and confident were you in identifying visual accessibility problems?

\end{enumerate}

\end{document}